\documentclass[english,aps,showpacs,twocolumn, superscriptaddress]{revtex4}
\usepackage[utf8x]{inputenc}
\usepackage{textcomp}
\usepackage{graphicx}
\usepackage{color}

\makeatletter

\@ifundefined{textcolor}{}
{%
 \definecolor{BLACK}{gray}{0}
 \definecolor{WHITE}{gray}{1}
 \definecolor{RED}{rgb}{1,0,0}
 \definecolor{GREEN}{rgb}{0,1,0}
 \definecolor{BLUE}{rgb}{0,0,1}
 \definecolor{CYAN}{cmyk}{1,0,0,0}
 \definecolor{MAGENTA}{cmyk}{0,1,0,0}
 \definecolor{YELLOW}{cmyk}{0,0,1,0}
}

\providecommand{\tabularnewline}{\\}

\makeatother

\usepackage{babel}
\begin{document}

\title{Adsorption geometry and the interface states: 

The relaxed and compressed~phases of NTCDA/Ag(111)}

\author{P. Jakob}
\email{peter.jakob@physik.uni-marburg.de}
\affiliation{Department of Physics, Philipps-Universität Marburg, 35032 Marburg,
Germany}

\author{N. L. Zaitsev}
\email{zaitsev@chemie.uni-marburg.de}
\affiliation{Department of Chemistry, Philipps-Universität Marburg, 35032 Marburg,
Germany}

\author{A. Namgalies}
\affiliation{Department of Physics, Philipps-Universität Marburg, 35032 Marburg,
Germany}

\author{R. Tonner}
\affiliation{Department of Chemistry, Philipps-Universität Marburg, 35032 Marburg,
Germany}

\author{I. A. Nechaev}
\affiliation{Centro de F\'{i}sica de Materiales CFM-MPC, Centro Mixto CSIC-UPV/EHU,
20018 San Sebasti\'{a}n, Spain}

\author{F.S. Tautz}
\affiliation{Peter Grünberg Institut, Forschungszentrum Jülich, 52425 Jülich,
Germany}

\author{U. Höfer}
\affiliation{Department of Physics, Philipps-Universität Marburg, 35032 Marburg,
Germany}

\author{D. S\'{a}nchez-Portal}
\affiliation{Centro de F\'{i}sica de Materiales CFM-MPC, Centro Mixto CSIC-UPV/EHU,
20018 San Sebasti\'{a}n, Spain}

\begin{abstract} 

The theoretical modelling of metal-organic interfaces represents a formidable challenge, especially in consideration of the delicate balance of various interaction mechanisms and the large size of involved 
molecular species. In the present study, the energies of interface states, 
which are known to display a high sensitivity to the adsorption geometry and electronic structure 
of the deposited molecular species, 
have been used to test the suitability and reliability of current theoretical approaches.
Two well-ordered overlayer structures (relaxed and compressed monolayer) of NTCDA on Ag(111) 
have been investigated using two-photon-photoemission to derive precise interface state energies for these closely related systems. 
The experimental values are reproduced by our DFT calculations using 
different treatments of dispersion interactions (optB88, PBE-D3) and 
basis set approaches (localized numerical atomic orbitals, plane waves) 
with  remarkable accuracy. 
Our results underline the trustworthiness, and some of the limitations, 
of current DFT based methods regarding 
the description of geometric and electronic properties of metal-organic interfaces.

\end{abstract}

\pacs{
73.20.-r,   
79.60.Dp     
78.47.-p,   
}

\maketitle

\subsection{Introduction}

Electronic devices based on organic thin films contain a number of functional units 
which are crucial for their proper operation. The 
interface to the metallic leads is one of them, as it constitutes the link between two electronic sub-units 
with very different properties; discrete molecular levels, on the one hand, 
meet the continuum band structure of an inorganic solid, on the other. 
The layer in direct contact to the substrate (contact primer layer) thereby 
plays a decisive role, determining the electronic structure at these interfaces, 
and it has been found that distinct interface states (IS) may form as a result of the 
nearby molecular layer \cite{Temirov_nature_2006,Schwalb08prl,Dyer_Persson_2010,Zaitsev2010,Marks11prb}.

Based on Scanning Tunnelling Microscopy (STM) and Spectroscopy (STS) data for 
PTCDA/Ag(111) \cite{Temirov_nature_2006} it was invoked that this electronic state, 
which is found to display free-electron-like dispersion, derives from an unoccupied molecular orbital (LUMO+1) 
which has broadened and shifted down in energy due to the interaction with the 
Ag(111) substrate \cite{Dyer_Persson_2010}. 
Dispersion then arises as a consequence of mixing with delocalized metallic states 
(substrate-mediated intermolecular coupling).
Alternatively, on the basis of the observed short life time of this interface state 
Schwalb et al. \cite{Schwalb08prl} argued 
that this state represents a modified surface state (SS) 
band that has been shifted upwards by the interaction with the molecular layer. 
Hybridization with unoccupied molecular orbitals may occur in addition, 
but this is not a prerequisite for a delocalized interface state to occur. 

Model calculations by Dyer and M.Persson \cite{Dyer_Persson_2010}, as well as Zaitsev et al. \cite{Zaitsev2010} 
support the 'upshifted surface state' model by Schwalb et al. and, for the particular system PTCDA/Ag(111),
confirm mixing with the LUMO+1 band of PTCDA, which has shifted down in energy upon contact with Ag(111). 
In particular, calculations in ref. \cite{Dyer_Persson_2010} reveal that outside the surface, where STM and STS 
are measured, the IS indeed exhibits a density distribution that clearly reflects its hybridization with the molecular orbitals.
Moreover, Marks et al. \cite{Marks11prb} find that, in agreement with the original conjecture 
of ref. \cite{Temirov_nature_2006}, the bending of the carboxylic oxygen atoms towards the Ag(111) surface 
enhances the molecular overlap with the up-shifted surface state.
Note that the nature of the IS as an upshifted SS quite naturally explains its free-electron-like dispersion. 
More detailed calculations \cite{Zaitsev12prb,Tsirkin15prb} have
been conducted to illuminate the dependence of the IS energy ($E_{IS}$) on the specific arrangement
and adsorption geometry (vertical bonding distance $d_{\perp}$, bending of the molecular framework, etc.) 
of the adsorbed species, the surface coverage, 
as well as to study the expected life times of the IS for various molecule-metal systems~\cite{Tsirkin15prb}.

The energy position of the interface state  $E_{IS}$ depends most sensitively on the vertical
bonding distance $d_{\perp}$ of the molecules. 
For the well studied PTCDA system $d_{\perp}$ decreases from 3.27~Å for 
PTCDA/Au(111) \cite{Henze07ss} to 2.86~Å for PTCDA/Ag(111) \cite{Hauschild_prl_2005,Bauer2012prb}
and to 2.81~Å for PTCDA/Ag(100) \cite{Bauer2012prb}. 
At the same time the upshift of the Shockley state $\Delta E_{IS}$ increases from 
0.16 eV \cite{Ziroff09ss} to 0.66 eV \cite{Schwalb08prl}, and to 0.95 eV \cite{Galbraith14jpcl}, respectively, 
resulting in an inverse scaling of 
${{\Delta E_{IS}} \over {\Delta d_{\perp}}}$ = -(1.2 - 1.7)~eV/Å $\approx$ -1.5~eV/Å \cite{Marks2014JElSpectr}. 
This high sensitivity of $E_{IS}$ on the adsorption geometry 
makes it an ideal model to test electronic structure calculations by comparing the results 
of different approaches and approximations with accurate experimental data.
Despite current improvements in the description of surface and interface properties of molecular layers, 
the correct description of interface state energies is by no means trivial.

Here, in a combined experimental and theoretical approach, we investigate the well-known
metal-organic model system NTCDA on Ag(111) to examine the dependence of interface state energies
on the molecule-substrate binding geometry.
NTCDA adsorbed on Ag(111) has been studied by various experimental techniques in the past  \cite{Stahl1998,Kilian2008,Stanzel2004,Stadler2007,Bendounan2007SS,Marks11prb,braatz_vibrational_2012,Tonner2016pccp,braatz_switching_2015}.
Two long range ordered monolayer phases (relaxed and compressed monolayers - denoted 
as r-ML and c-ML in the following) have been identified \cite{Stahl1998,Kilian2008,braatz_vibrational_2012,braatz_switching_2015}. 
A moderate interaction strength of NTCDA with Ag(111) 
has been concluded based on the observed downward shift of molecular orbitals upon 
adsorption \cite{Bendounan2007SS}; vibrational frequency shifts are particularly strong for the carboxyl group 
located at the corners of the molecule, which has been attributed to a chemical interaction with the silver substrate, in 
addition to the ubiquitous van der Waals (vdW) interactions \cite{braatz_vibrational_2012,Tonner2016pccp}.
 
Using the normal incidence x-ray standing wave technique (NIXSW), $d_{\perp}$ values 
have been determined for the relaxed and the compressed NTCDA monolayer phases on 
Ag(111) \cite{Stanzel2004,Stadler2007}. 
The molecular arrangements of both long range ordered overlayers have 
only very recently been resolved on the basis of low temperature STM data \cite{braatz_switching_2015}.
For r-ML the parallel oriented NTCDA molecules are arranged in a brickwall-type fashion and 
two inequivalent molecules within the rectangular unit cell can be distinguished. 
Unlike r-ML with all NTCDA aligned identically (long molecular axis oriented along 
Ag(111) atom rows), the c-ML phase comprises two different azimuthal 
orientations of NTCDA. More precisely, the flat lying molecules are forming a 
herringbone structure with four inequivalent NTCDA within the unit cell. 

In this context, we stress that the structural details regarding the lateral ordering are crucial 
to derive meaningful quantities for the vertical bonding distances, 
molecular deformations, adsorption energies and electronic levels of the 
molecular layer from calculations, and thereby allow us to establish a correlation 
between $E_{IS}$ and $d_{\perp}$. 

The intriguing role of NTCDA/Ag(111) as a model system not only refers to the comprehensive knowledge base 
regarding its electronic and structural properties, but additionally relies on the fact that two layers with 
almost identical composition (coverage, molecular orientation) exist. 
Specifically, the two long range ordered NTCDA monolayer phases differ in their 
coverages by only 10\% ($\Theta_{NTCDA}$ = 1.0 monolayers (ML) for c-ML, versus 0.9 ML for r-ML). 
As we will show here, these two phases exhibit interface states at well-defined 
but distinctly different positions, with the r-ML IS located at a higher energy. 
This counter-intuitive result seems to stem 
from the balance between the effect of the molecular adsorption distance and the different coverages. 
Therefore, NTCDA/Ag(111) represents an ideal playground to gauge the ability of current 
theoretical methods based on 
density-functional theory (DFT) to predict the molecular binding geometries and the relation 
between those geometries and $E_{IS}$. This is particular important since, in general, 
to obtain valid structural information of the molecular adsorption configuration represents 
an experimental challenge and the adsorption height is known from experiments only for a few
selected systems (sometimes with sizeable error bars).
For NTCDA/Ag(111) the accumulated experimental knowledge in the literature, plus the data presented here
for the surface electronic structure using two-photon-photoemission (2PPE) 
measurements, pose a well-defined and stringent test to the predictive capabilities
of existing DFT  methods. 
Eventually, theory may provide trustworthy structures (via dispersion corrected DFT) and
reliable $E_{IS}$ even in cases where e.g. the experimental adsorption
height is not known.

\subsection{Experimental Methods}

The 2PPE experiments and the preparation of the sample was performed in the same ultra high vacuum (UHV) chamber 
at a base pressure below $1 \times10^{-10}$~mbar. 
The Ag(111) crystal was cleaned by Ar-sputtering cycles (3~$\mu$A, 700~V, 15~min. 373~K) and 
subsequent annealing (773~K, 5~min).
The NTCDA layers were grown at a sample temperature of 90~K and a rate of 0.4~ML per minute. 
After deposition of $\approx$1.5~ML, we prepared the c-ML and r-ML phases according to the recipe of 
Braatz et al. \cite{braatz_vibrational_2012}, i.e., by thermal annealing to
either 360 - 370~K (c-ML) or 400~K (r-ML) and using a temperature ramp $\approx 1$\ K/s. 
The long range order and purity of the film was checked by low energy electron 
diffraction (LEED) and X-ray photo-electron spectroscopy (XPS).

We used the laser setup described by Sachs et al. \cite{Sachs09jcp} (PE-I) with a pump laser energy 
of $\hbar \omega _{\rm pump}=3.08$\ eV, (pulse width 68~fs) and a probe energy of 
$\hbar \omega _{\rm probe}=4.62$\ eV, (pulse width 77~fs). The incident beams were focused onto the sample 
at an angle of 75$^{\circ}$ with respect to the surface normal; 
the spectra reported here refer to a sample temperature of 90 K and were recorded in normal emission, 
i.e., at the $\Gamma$-point. 
Special care has been taken to avoid space-charge effects, which might adversely influence 
the precisely measured energetic positions. 
The UV beam photon energy, which was very close to the work function of the clean sample, 
has been reduced to a pulse energy below 0.05~nJ to avoid detrimental effects due to a 
low-energy one-photon background. The time delay has been set to zero to maximize signals. 

The prominent Shockley surface state of the Ag(111) surface has been used to calibrate the
energy scale of our experimental setup.
According to Reinert et al. \cite{Reinert01prb} it has a binding energy of 63 meV at 30 K ($\Gamma$-point).
Panagio et al. derived a somewhat less precise value of 26 meV at 300 K and, 
more importantly, they examined the temperature dependence of this state 
in the temperature range 50 - 550 K \cite{Paniago95ss1}. 
Using the precise low T value of Reinert et al., and applying the observed T-dependency 
(Figure 5b of ref. \cite{Paniago95ss1}), the surface state for clean Ag(111) is expected 
at 59 meV below the Fermi level at our sample temperature of 90 K.

\subsection{Calculation details}

First-principles electronic structure calculations were performed
in the framework of density functional theory (DFT) using the SIESTA code~\cite{SIESTA1,SIESTA2}, 
with localized numerical atomic orbitals as a basis set, and the VASP 
code~\cite{kresse_efficiency_1996}, where a plane-wave expansion is used to represent the 
electronic wave-functions. In the SIESTA code core electrons 
are replaced by norm-conserving Troullier-Martins type pseudopotentials~\cite{TMpseudos},
while in VASP the so-called projector-augmented wave (PAW) method is applied
\cite{Bloechl1994,Kresse_Joubert_1999}. 

Since local and semilocal density functionals are known to provide a poor description of
the adsorption geometry of metal-organic interfaces, here we use 
exchange-correlation (\textit{xc}) functionals that explicitly account for long range, non-local 
dispersion interactions (vdW-DF2).
In particular, we chose the optB88 functional by 
Klimes et al.~\cite{klimes_chemical_2010,klimes_van_2011},
providing the best equilibrium lattice constant and bulk modulus for bulk silver 
(respectively, 4.172~\AA\  and 119~GPa in our SIESTA calculations) among the vdW-DF2 functionals we tested. 
Additionally we performed calculations within the PBE-D3(BJ) scheme~\cite{grimme_consistent_2010,grimme_2011}, 
that incorporates an empirical correction to include dispersion forces 
on top of the generalized gradient approximation PBE functional~\cite{PBE}.

A scheme of periodically repeated slabs was used to describe the Ag(111) surface. 
To avoid direct interaction between the periodic images of the system 
a vacuum layer of $\sim$11~\AA\ was included in the direction perpendicular to the surface ($z$-direction). 
In order to get the final geometric structures, we allowed full relaxation of the NTCDA molecular monolayer 
and the two underlying Ag layers of the four-layer silver slab until the maximum force becomes less than 0.01 eV/\AA. 
For the calculation of interface states we took slabs comprising 12 layers. 
According to our tests, with this increased number of layers, the energy of the IS is converged within 10 meV 
with respect to the slab thickness. 
Notice that both the SS and the IS are located within the projected band gap of Ag(111), 
but still rather close to the bulk bands and, thus, extend far into the substrate, 
requiring the use of rather thick slabs.
The surface Brillouin zone was sampled using a Monkhorst-Pack scheme 
with 8 and 6 k-points for the relaxed and compressed phase, respectively.

For SIESTA calculations, we used a double-$\zeta$ polarized (DZP) basis generated
within the soft-confinement scheme. Silver orbital radii corresponded to an 
{\it energy shift}~\cite{SIESTA2} of 180~meV, providing 3.72~\AA\ for the 5$s$ orbital. 
For a better description of the Ag(111) surface~\cite{garcia-gil_optimal_2009} we used an enlarged range of Ag pseudoatomic orbitals for atoms in the surface (uppermost and lowermost) layers of the slab. 
We explicitly checked that the position of the Ag(111) surface state ceases changing when the radius 
of the 5$s$ orbitals of Ag reaches a value of 5.15~\AA, which corresponds to 10~meV of {\it energy shift}. 
The same {\it energy shift} value was used to produce the basis functions of all molecular atoms.
Note that usage of extended orbitals for all the Ag atoms in the slab results in 
an insignificant change of the electronic properties.  

A uniform mesh for numerical integration and solution of the Poisson
equation was specified by an energy cutoff of 250 Ry; for structure
optimization (to avoid numerical noise in the forces) this value was increased
to 400 Ry. For VASP calculations, the plane-wave cut-off was fixed at 350~eV.

\begin{figure}[h]
\includegraphics[width=0.9\columnwidth]{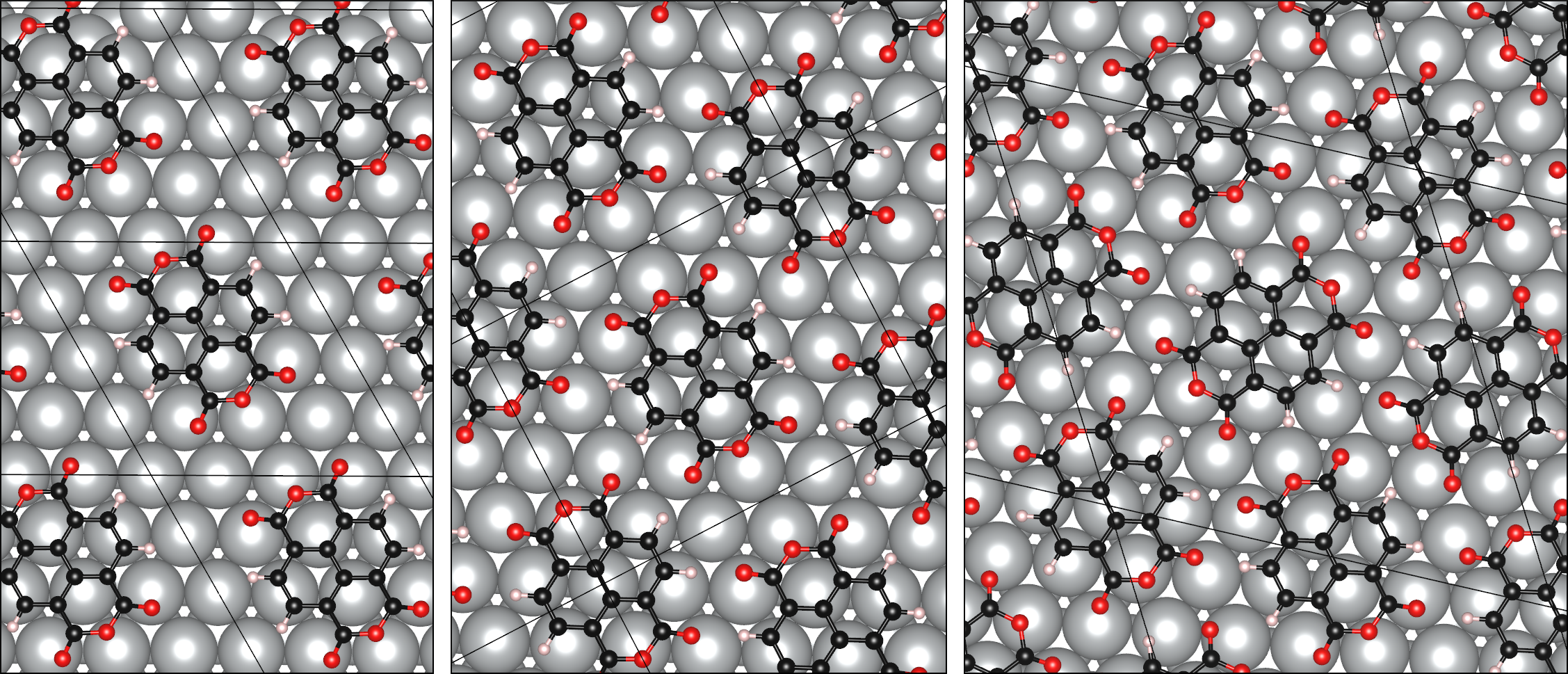}

\caption{\label{fig:start-geo} (Color online) The surface unit cells and initial arrangements
of NTCDA molecules on Ag(111). The hypothetical 4x4 commensurate structure (left) which comprises 
identical adsorption geometries for all NTCDA and which is characterized by negligible intermolecular interactions 
has been analyzed recently by Tonner, Rosenow, and Jakob \cite{Tonner2016pccp}.
The arrangement of NTCDA molecules in the experimentally observed relaxed (middle) and compressed (right) phases 
has been derived from low T STM data \cite{braatz_switching_2015}.}
\end{figure}

The unit cells and arrangements of the molecules within them are depicted
in Figure \ref{fig:start-geo}. The artificially diluted (4x4) structure
contains one molecule per unit cell which is located above bridge site positions
of the silver substrate (molecule center), and with the central C-C bond aligned 
along Ag atom rows (Fig. \ref{fig:start-geo} left). There are two molecules in the
unit cell of r-ML. One molecule occupies a bridge site, the other one the 
on-top position. The unit cell of c-ML contains four molecules, 
two of them occupy the same adsorption sites as for the relaxed monolayer phase. 
The remaining two molecules are rotated by about $90{}^{\circ}$
degrees with respect to first ones and sit on bridge sites. The arrangements
of the molecules in the r-ML and c-ML phases were deduced from 
high resolution STM images \cite{braatz_switching_2015}. The analyzed structures
refer to different coverages 0.672 ML, 0.896 ML and 1 ML for the 
dilute (4$\times$ 4), r-ML and c-ML phases, respectively; one ML is thereby defined 
as the maximum amount of parallel oriented NTCDA, which can be accomodated 
on Ag(111), i.e. for the compressed monolayer.

\subsection{Results}

\subsubsection{Experimental Results}

\begin{figure}[!]
\includegraphics[width=1.0\columnwidth]{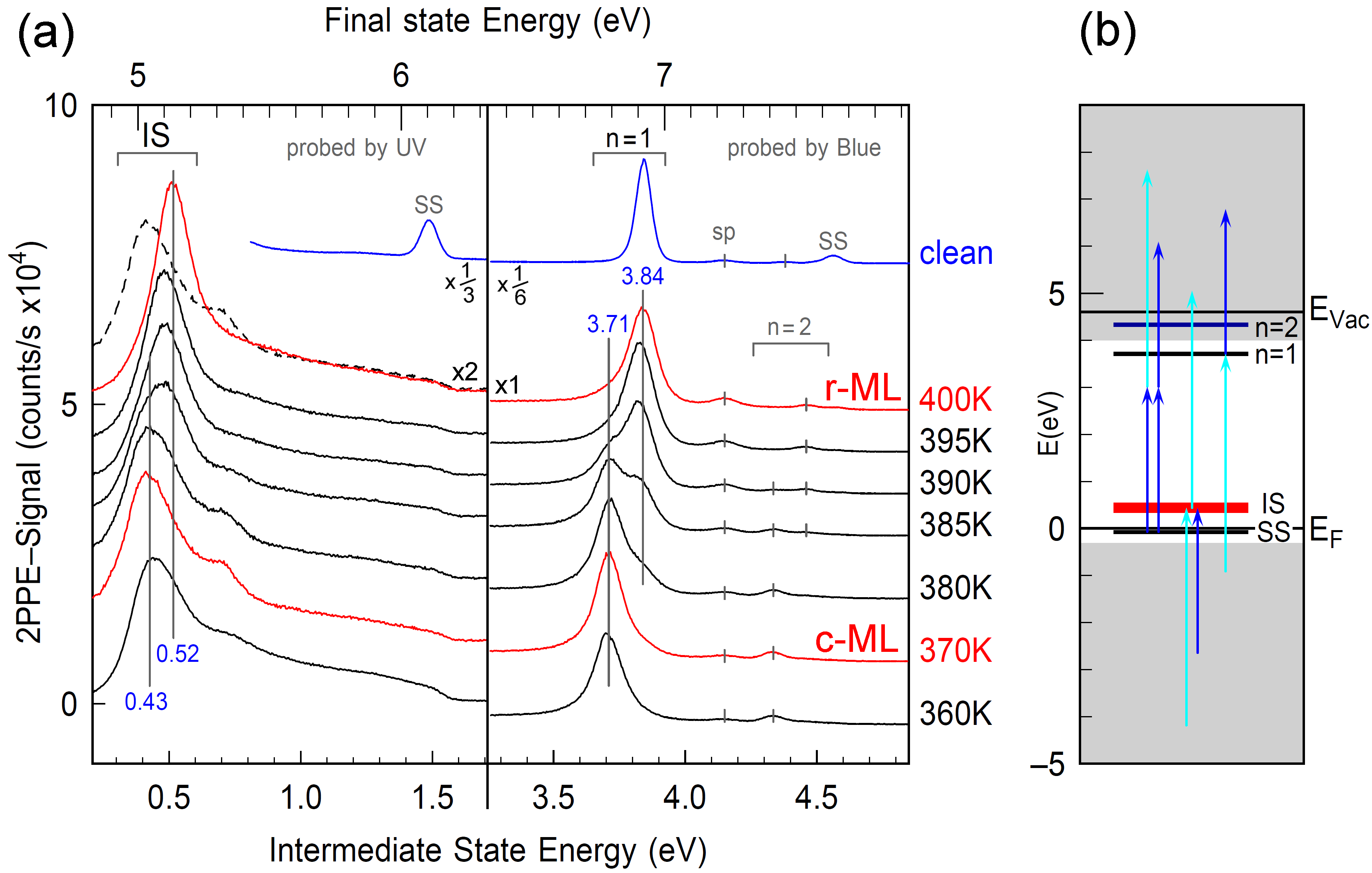}

\protect\caption{\label{fig:cML_to_rML_series} 
(Color online) (a) Series of 2PPE spectra of NTCDA on Ag(111) in the coverage range 0.9 - 1.0 ML. 
Starting with the compressed monolayer ($\Theta_{NTCDA}$ = 1.0 ML) the NTCDA 
coverage has been successively reduced by stepwise annealing until a complete 
relaxed monolayer ($\Theta_{NTCDA}$ = 0.9 ML)  is obtained. 
The respective annealing temperatures are indicated to the right of the right panel.
Note that the vertical scale displaying the NTCDA/Ag(111) interface state (left panel) has been
enlarged by a factor of 2 with respect to the spectral region of image potential states (right panel).
In order to aid direct comparison of the interface states for r-ML and c-ML 
the latter trace (dotted line) has been overlaid to the r-ML curve. 
The spectrum of the clean Ag(111) surface (topmost curve) covering the Shockley surface state (SS), 
the Ag-bulk sp band (sp), and the image potential states (n = 1, n = 2) has been added for comparison.
Final state energies of the emitted electrons, as well as the energies of the intermediate state 
(not applicable to SS, see main text) are given with respect to the Fermi energy ($E_F$). 
(b) Excitation and probing scheme indicating the various spectral features of the compressed monolayer. 
The length of the blue and turquoise arrows is proportional to the photon energy of the blue and UV laser beam. 
The grey areas display the projected Ag(111) bulk bands at the $\Gamma$-point. 
 }
\end{figure}

In  Figure \ref{fig:cML_to_rML_series}, 2PPE spectra of c-ML and r-ML are displayed. 
The NTCDA/Ag(111) sample has been prepared by NTCDA deposition at low T $\approx$ 80 K 
(for details, see the experimental section) and annealing to successively higher temperatures 
(and recooling thereafter for data collection); 
due to desorption of NTCDA in the 
temperature range 370 - 400 K the complete monolayer (c-ML phase) gradually
transforms to the less dense r-ML phase. 
The image potential states of the NTCDA monolayer thereby serve as an indicator 
for the quality and integrity of the layers. Specifically, discrete bands associated 
with the two phases are observed at distinctly different energies. 
The n = 1, 2 image potential states of c-ML are located at 3.71 eV and 4.33 eV, 
and those of r-ML at 3.84 eV and 4.46 eV above $E_F$, respectively 
(see Fig. \ref{fig:cML_to_rML_series}). 
The spectrum of the clean silver is added on top of the figure for comparison. 
In contrast to the emission of the unpopulated interface and image potential states, the populated Shockley surface state (SS), 
which was used for the energetic calibration, can only be emitted by a two photon process. 
It appears at the final state energy positions of 6.1~eV or 7.64~eV, depending on whether the emission is induced by 
absorption of two blue photons $\hbar \omega _{\rm blue+blue}=6.16$\ eV, 
or one blue and one UV photon $\hbar \omega _{\rm blue+UV}=7.70$\ eV, respectively.

The absence of any unwanted sidebands or shoulders for n = 1, 
as well as the absence of the prominent Ag(111) surface state are taken 
as proof for the excellent quality and homogeneity of the prepared phases. 
Moreover, their narrow line shapes support the idea that this transition proceeds by a 
gradual shift of the boundary separating both phases as has been suggested by 
Stahl et al.  \cite{Stahl1998} based on STM data.

In the region of the interface state, i.e. 0-1 eV above $E_F$ we identify distinct 
spectral features at 0.43 eV (c-ML) and 0.52 eV (r-ML). 
For both bands the read-out proceeds exclusively via UV absorption;  
excitation of electrons in this unoccupied electronic level may be achieved by means of either blue or UV light. 

Comparison of the IS energies for c-ML and r-ML thus reveals a notably lower value for 
c-ML, which may indicate a larger vertical bonding distance of NTCDA for this phase 
as compared to r-ML. Since the NTCDA coverage is about 10\% less for r-ML, an 
extrapolation to coverages equal to the c-ML phase would cause an extra upshift of the r-ML interface state energy, 
implying an accordingly larger difference in vertical bonding distance between the c-ML and r-ML phases. 

We note that for less dense layers ($\Theta < \Theta_{max}$) empty spaces between molecules 
cause leakage of the IS wave function into the vacuum, leading to a lowering of $E_{IS}$. 
The effect of an incomplete molecular layer has been examined theoretically for an idealized, planar layer 
of NTCDA on Ag(111) and will be discussed in more detail below. 
An estimate based on these findings (extrapolation to full saturation of the monolayer) would 
produce a 20 - 40 meV upshift of the interface state, assuming a vertical bonding distance of 2.9 - 3.0~\AA. 
According to the projected shift of $E_{IS}$ versus vertical bonding distance 
of about 1.5 eV/\AA\ \cite{Marks2014JElSpectr}, the 
observed (90 meV), plus extrapolated (20-40 meV), difference in interface state energy of~$+$(110 - 130)~meV 
for r-ML with respect to c-ML would translate into a $0.08 \pm 0.01$~\AA\ closer
average height of NTCDA above Ag(111) for r-ML as compared to c-ML. 

This value may be compared to NIXSW literature data which have been obtained 
for both ordered phases of NTCDA/Ag(111) \cite{Stanzel2004, Stadler2007}. 
For r-ML and c-ML,  $C_{1s}$ and  $O_{1s}$ data have been analyzed to derive adsorption heights 
(averaged over all C-atoms) of 3.02±0.02 Å for r-ML and 3.09±0.04 Å for c-ML \cite{Stanzel2004}. 
Later on more precise values have been derived for r-ML (see Table 1).
These values are notably lower than expected for pure van der Waals interacting 
molecular species and underline the significance of chemical interactions for NTCDA on Ag(111). 
This finding is also in accordance with a marked chemical shift in XPS binding energies 
\cite{Schoell2004JPCB} and the partial filling of the former LUMO of adsorbed NTCDA \cite{Bendounan2007SS}. 
According to NIXSW findings for  $C_{1s}$ and  $O_{1s}$ core levels the planar molecular 
structure of NTCDA becomes warped when adsorbed on Ag(111). This has been 
attributed to covalent bonding of NTCDA to substrate atoms via the four carboxyl groups, 
in accordance with similar findings for PTCDA on Ag(111) \cite{Hauschild_prl_2005}.
Such covalent bonding and the accompanying downward bending of carboxyl oxygens agree 
favorably with the pronounced redshift of respective C=O stretching modes for 
NTCDA on Ag(111) \cite{braatz_vibrational_2012,Tonner2016pccp,braatz_switching_2015}.

In NIXSW data,  as well as in our theoretical calculations, the extrapolated bulk positions of 
Ag(111) surface atoms have been used as the reference frame ($z_{ref} = 0$). 
True spacings between molecule and metal then 
additionally comprise surface relaxation effects which may add up to values of about 0.1~Å and 
additionally depend crucially on the type of adsorbed species. 
For NTCDA/Ag(111) the difference between the extrapolated bulk position of Ag(111) 
and the calculated average position of Ag surface atoms amounts to 0.02~\AA\ at most 
for all structures considered in this study.

Analysis of adsorbed PTCDA or NTCDA on Ag(111) \cite{Bauer2012prb, Tonner2016pccp}
provided clear evidence of a non-negligible relaxation of Ag surface atoms which did 
comprise not only an overall contraction of the vertical bonding distance but also introduced 
an extra corrugation of surface atoms underneath of PTCDA or NTCDA. 
As shown below, our calculations confirm that this is the case also for the NTCDA c-ML and r-ML
phases discussed here. 

To conclude this section, our observed IS energies for c-ML and r-ML are fully consistent 
with existing NIXSW data. The next step is therefore to use DFT to derive a model of 
both parallel oriented NTCDA/Ag(111) phases including their detailed adsorption 
geometries and, moreover, to determine the corresponding molecular electronic structure 
and interface state energies.

\subsubsection{Theoretical Results}

We found optimized geometries for the r-ML and c-ML phases of NTCDA on Ag(111) 
by means of VASP (plane waves) and SIESTA (numerical pseudoatomic orbitals)
calculations performed with different schemes to take into account
vdW interactions, the optB88 functional and the PBE functional plus D3 empirical dispersion corrections.
Although these approximations provide slightly different
bulk properties (e.g. the optB88 functional overestimates the equilibrium lattice constant) 
the final adsorption geometries are quite similar (Table \ref{tab:Rn-Vertical-adsorption-distance}).

The calculated optimal structures of the relaxed phase are in good agreement 
with experimental measurements (see Table \ref{tab:Rn-Vertical-adsorption-distance}). 
In accordance with NIXSW data \cite{Stanzel2004, Stadler2007} we find that the $O_{carb}$ position 
is shifted down with respect to the carbon core, while $O_{anh}$ resides at the same plane 
with $C_{func}$. 
The vertical bonding distance of the C backbone for the bridge molecule is 0.02 Å higher
than for the one at on-top sites. 
The (vertical) distortion along the long axis of NTCDA, i.e. $\delta (C_{core}-C_{func})$, 
is about 0.15~Å, while the distortion along the short axis, i.e $\delta (C_{core}-H)$, 
amounts to 0.1~Å for both type of molecules. 
In general, the overall bending $\delta (C_{core}-O_{carb})$ of the on-top molecule 
is somewhat smaller than that of bridge bonded NTCDA by 0.1~Å (see Table \ref{tab:Rn-Vertical-adsorption-distance}), 
and the respective values seem largely independent of the choice of the \emph{xc}-functional. 
As we are not sure how the averaging of individual $C_{1s}$ signals is contributing to the experimental 
values of carbon vertical distances $d_{\perp}$ we should not be overly concerned regarding residual differences, 
since the overall agreement is very good. 

\begin{table*}[tb]
\caption{\label{tab:Rn-Vertical-adsorption-distance} 
Vertical adsorption distances (values given in~\AA ~and with respect to the 
extrapolated bulk positions of the Ag(111) surface atoms) for the relaxed NTCDA/Ag(111) monolayer phase. 
The bending of the molecules is quantified by $\delta (C_{core} - O_{carb})$. 
The different parts of the molecule are specified with the following notation: $C_{c}$ --- two central carbon atoms;  
$C_{core}$ --- four inner carbon atoms of naphtalene core; $C_{H}$ --- four carbon atoms connected to H-atoms; 
$C_{func}$ --- four carbon atoms connected to carboxylic oxygens $O_{carb}$; $O_{anh}$ --- anhydride oxygens. 
The computed lattice parameters $a_{Ag}$ of Ag using each computational scheme are also indicated. 
The experimental value for C-atoms corresponds to averaging over all C-atoms for r - ML.
}

\begin{ruledtabular}

\begin{tabular}{lcccccccccc}
 & \multicolumn{2}{c}{Siesta: optB88} &  & \multicolumn{2}{c}{Vasp: optB88} &  & \multicolumn{2}{c}{Vasp: PBE-D3 } &  & exp\footnote{NIXSW\cite{Stadler2007}} \tabularnewline
 & \multicolumn{2}{c}{$a_{Ag}=4.17$~\AA} &  & \multicolumn{2}{c}{$a_{Ag}=4.14$~\AA} &  & \multicolumn{2}{c}{$a_{Ag}=4.07$~\AA} &  & $a_{Ag}=4.06$~\AA \footnote{Experimental value corrected for zero-point energy effects \cite{csonka_assessing_2009}}\tabularnewline
\hline 
 & top & bridge &  & top & bridge &  & top & bridge &  & \tabularnewline
\cline{2-3} \cline{5-6} \cline{8-9} 
$C_{c}$ & 3.06 & 3.09 &  & 3.11 & 3.12 &  & 3.05 & 3.10 &  & \tabularnewline
$C_{core}$ & 3.01 & 3.03 &  & 3.06 & 3.06 &  & 3.00 & 3.04 &  & \textbf{2.997}\tabularnewline
$C_{H}$ & 2.94 & 2.96 &  & 3.02 & 3.01 &  & 2.93 & 2.96 &  & \tabularnewline
$C_{func}$ & 2.86 & 2.87 &  & 2.92 & 2.89 &  & 2.88 & 2.88 &  & \tabularnewline
$O_{carb}$ & 2.69 & 2.61 &  & 2.73 & 2.64 &  & 2.69 & 2.63 &  & \textbf{2.747}\tabularnewline
$O_{anh}$ & 2.85 & 2.85 &  & 2.92 & 2.89 &  & 2.91 & 2.90 &  & \textbf{3.004}\tabularnewline
$H$ & 2.91 & 2.93 &  & 3.0 & 2.99 &  & 2.91 & 2.94 &  & \tabularnewline
$\delta (C_{core} - O_{carb})$ & \emph{0.32} & \emph{0.42} &  & \emph{0.33} & \emph{0.42} &  & \emph{0.31} & \emph{0.41} &  & \tabularnewline
\end{tabular}

\end{ruledtabular}
\end{table*}

\begin{figure}[tb]
\includegraphics[width=1.0\columnwidth]{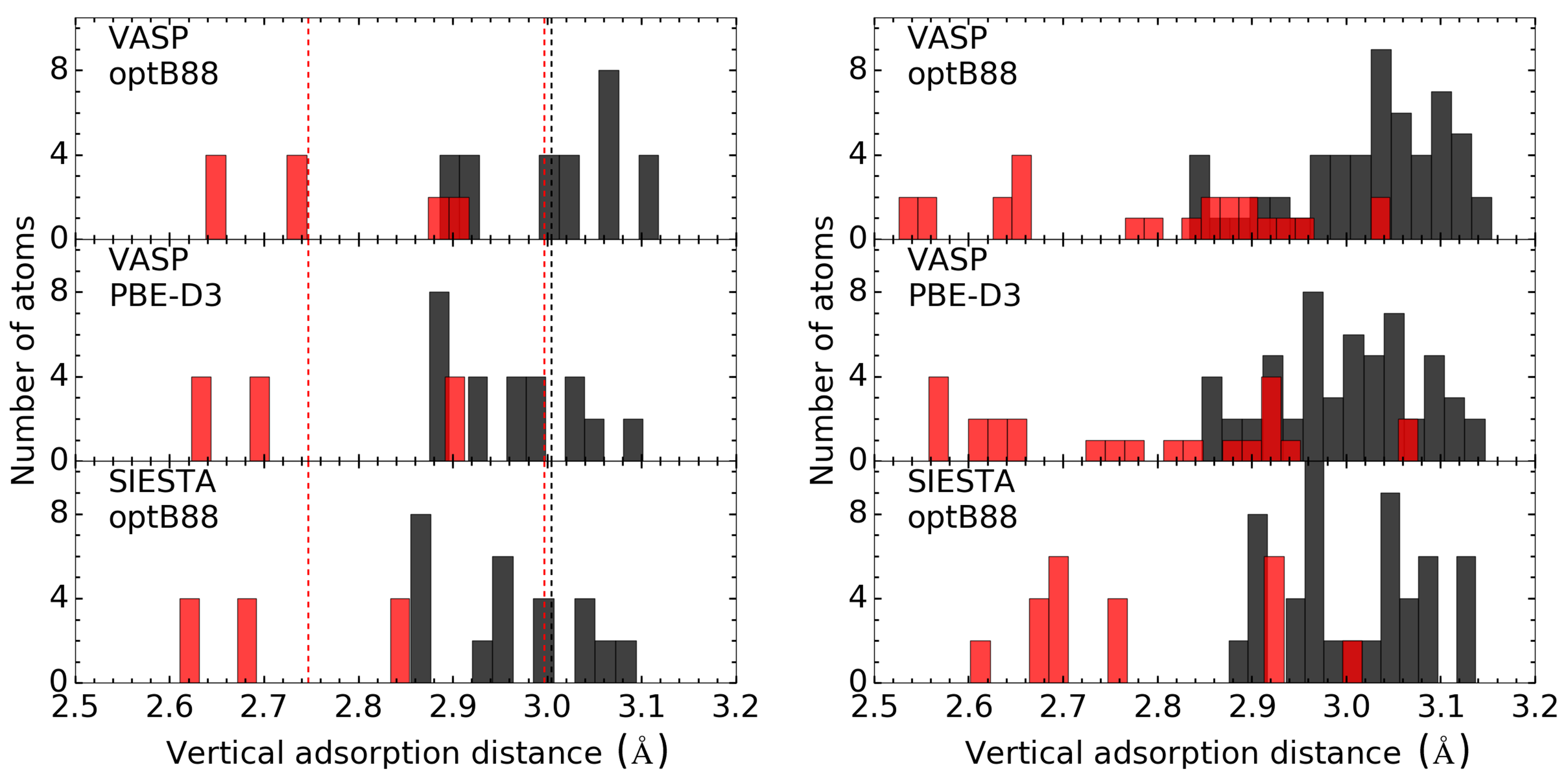}

\caption{\label{fig:dz-distribution} (Color online) The distribution of vertical adsorption
distances in equilibrium geometry for the relaxed (left) and compressed (right) phases. 
The oxygen and carbon atoms are represented by red and black bins, respectively. 
According to Stanzel et. al. \cite{Stanzel2004} vertical
heights of carbon atoms in the relaxed and the compressed phase of NTCDA on Ag(111) 
are 3.02 and 3.09~\AA, respectively. More recent measurements for the relaxed phase \cite{Stadler2007} 
are depicted as vertical dashed lines.}
\end{figure}

\begin{figure}
\includegraphics[width=0.8\columnwidth]{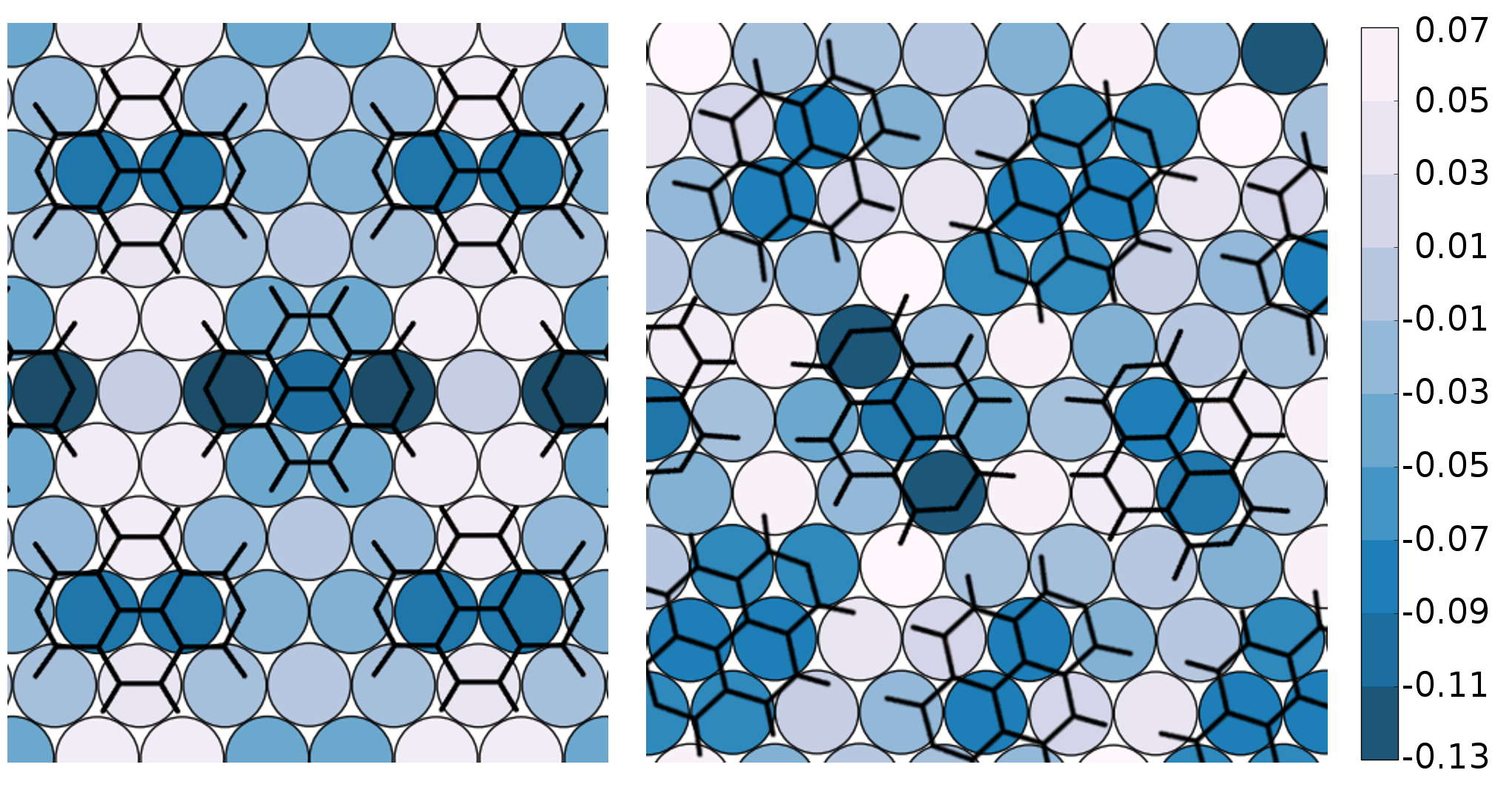} 

\caption{\label{fig:The-vertical-deviation} (Color online) The vertical deviation (in Å) of
the silver atoms in the topmost layer from their idealized positions 
(extrapolated bulk positions of Ag(111) surface atoms).
The left and right panels show the arrangement of molecules in the
relaxed and the compressed NTCDA/Ag(111) monolayers, respectively. The calculation was
performed using SIESTA and the optB88 functional. 
Positive (negative) values refer to displacements towards the vacuum (bulk). 
}
\end{figure}

A graphical illustration of vertical positions of the various functional groups 
of the NTCDA molecule within the relaxed and the compressed phase is 
provided in Figure \ref{fig:dz-distribution}. 
Again the three different theoretical approaches (see Table \ref{tab:Rn-Vertical-adsorption-distance}) 
have been applied.
In accordance with available experimental data \cite{Stanzel2004}, the vertical height
of the carbon core is slightly larger ($\sim$ 0.05~\AA) for c-ML as compared to r-ML. 
In general, our calculations, however, display similar adsorption geometries for the two phases 
of NTCDA/Ag(111) (Fig. \ref{fig:dz-distribution}).
This in particular applies to the aligned bridge and on-top adsorbed NTCDA molecules of c - ML, 
i.e. those with identical coordination to the Ag(111) surface atoms as in the r-ML phase; 
interestingly, even the silver atoms underneath these NTCDA are subject to virtually the same 
modification (see Fig. \ref{fig:The-vertical-deviation}). 
Within the c-ML the bridge and on-top molecules are slightly rotated by $2^{\circ}$ with respect
to the molecules of the relaxed monolayer phase (which are strictly aligned along the Ag substrate atoms); 
in fact, our finding of a lower total energy for a slightly rotated molecular arrangement is in accordance 
with the local symmetry ($C_2$) experienced by the individual NTCDA within the c-ML molecular layer. 
This is actually the reason why the bridge bonded molecules become slightly twisted
with respect to their diagonal. The remaining two inequivalent molecules within the c-ML phase, 
one of which lies at the center and another one at the edge of the drawn unit cell (Fig. \ref{fig:start-geo}), 
they both occupy bridge sites, however, with a strong azimuthal misorientation with respect to Ag(111) atom rows, 
and which leads to a twisting of their molecular planes as well.

The molecule-substrate interaction causes not only a distortion of the molecules 
but, additionally, a noticeable corrugation of the silver surface layer, 
especially for the on-top molecule. The attractive interaction of carboxylic
oxygens with underneath silver atoms pulls them out from their ideal
plane. Carbon atoms on the longitudinal axis and anhydride oxygens, 
on the other hand, push the underlying silver atoms downwards towards the bulk 
(Fig. \ref{fig:The-vertical-deviation}), indicative for a repulsion between them. 
In case of the bridge molecule we observe a similar trend but with smaller displacements. 
Thus, the overall interaction between the molecules and the
substrate comprises two major contributions, similar to the case of PTCDA monolayers
on different Ag surfaces \cite{Bauer2012prb}: Specifically, the derived 
geometry is determined by the interplay between attraction of the
functional groups with silver atoms, repulsion of the carbon backbone
from the substrate and the ability of the molecule and the surface to accommodate such distortions.

The surface and interface state energies ($E_{SS}$ and $E_{IS}$) for r-ML and c-ML phases 
are presented in Table \ref{tab:The-interface-states}. 
It is apparent that the absolute position of the interface states and the difference $\Delta E$ between
the energy of SS and IS is directly dependent on the lattice parameter of the
substrate. Bigger lattice constants provide higher absolute positions of SS and IS
and somewhat smaller values of $\Delta E$ (Table \ref{tab:The-interface-states}). 
Calculations performed with the same \emph{xc}-functional, but with various choices
of basis function types (and codes) give similar values for $\Delta E$.
In general, it is difficult to compute the absolute positions of IS and SS accurately 
(slabs containing a very large number of atomic layers are required), 
or to determine the $E_F$ with high precision (very dense k-sampling required 
in combination with the large supercells used here), because very large computational resources are needed, 
especially in the case of plane wave calculations. 

For a meaningful description and comparison of the two NTCDA phases  
the applied methods of calculation should use the same parameters. 
Moreover, the importance of using $E_{SS}$ as a reference scale for $E_{IS}$ is apparent from 
Table \ref{tab:The-interface-states}: Values of $E_{IS}$ are found in better agreement with the experiment in this case. 
The improved accuracy of thus obtained interface state energies 
is attributed to a compensation of (calculation) errors in deriving $E_{SS}$ and $E_{IS}$ values
and the well known energetic position (experiment) of the Ag(111) surface state  \cite{Reinert01prb,Paniago95ss1}.

\begin{table*}[tb]
\caption{\label{tab:The-interface-states} The surface state (SS) and interface
state (IS) energies for the layers displayed in Figure \ref{fig:start-geo}  
and calculated with different approaches. All energy values are given in meV.}

\begin{ruledtabular}
\begin{tabular}{ccccccccccccc}
 & {XC}
 & $E_{IS}$ & $E_{SS}$ & $\Delta E$ &
 & $E_{IS}$ & $E_{SS}$ & $\Delta E$ &
 & $E_{IS}$ & $E_{SS}$ & $\Delta E$\tabularnewline
\cline{2-13} 
 &
 & \multicolumn{3}{c}{4$\times$4} & 
 & \multicolumn{3}{c}{relaxed ML} & 
 & \multicolumn{3}{c}{compressed ML}\tabularnewline
\cline{3-5} \cline{7-9} \cline{11-13} 
VASP
 & optB88 (
 $a_{Ag}=4.14$~Å )
 &  &  &  &
 & 433 & -72 & \emph{505} &
 & 481 & -54 & \emph{535}\tabularnewline
 & PBE-D3 ($a_{Ag}=4.07$~Å)
 &  &  &  &
 & 424 & -164 & \emph{588} &
 & 450 & -157 & \emph{607}\tabularnewline
Siesta
 & optB88 ($a_{Ag}=4.17$~Å)
 &  445 & -8  &
 \emph{453}  &
 & 540 & 0 & \emph{540} &
 & 554 & -5 & \emph{559}\tabularnewline
 & PBE-D3 ($a_{Ag}=4.07$~Å)
 &  &  &  &
 & 468 & -137 & \emph{605} &
 & 480 & -153 & \emph{633}\tabularnewline
\end{tabular}


\end{ruledtabular}
\end{table*}

\begin{figure}[t!]
\includegraphics[width=1.0\columnwidth]{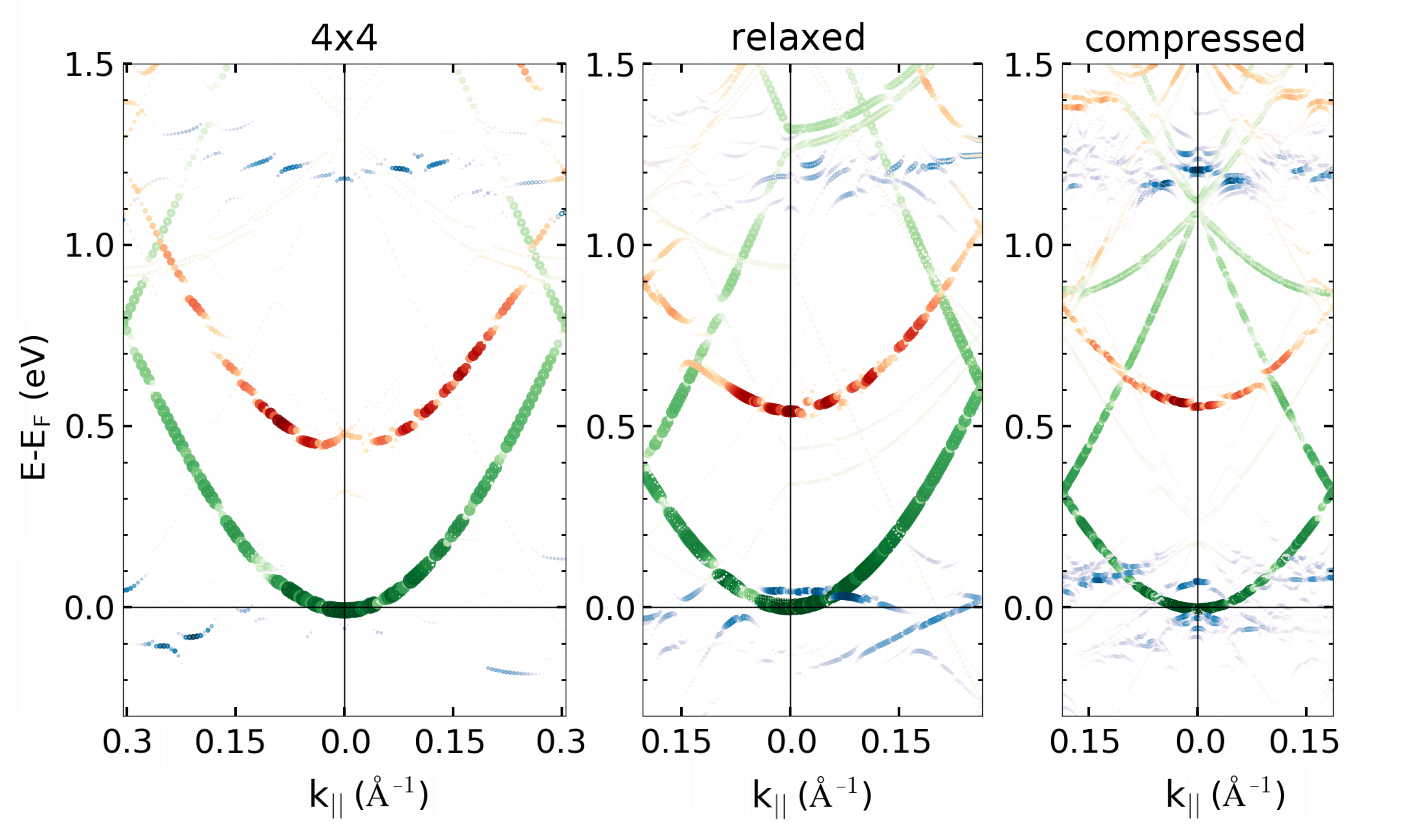}

\caption{\label{fig:Pbands} 
(Color online) Band structure  of the different NTCDA/Ag(111) structures studied here 
using the SIESTA code, the optB88 \emph{xc}-functional and a slab containing 
12 layers to represent the silver substrate. The colors and intensity of the 
symbols indicate the weight on selected atomic orbitals. 
Red symbols indicate the contribution from \emph{$p_{z}$} orbitals of the topmost silver layer 
and reveal the interface state. 
Green symbols highlight \emph{$p_{z}$} character on the silver layer from the clean side of the slab
and reveal the surface state. Carbon and oxygen \emph{$p_{z}$} character
are shown in blue.}
\end{figure}

The computed values for $\Delta E$ in Table \ref{tab:The-interface-states}
must be compared with our experimental value (assuming $E_{SS}\simeq$~-59~meV at T = 90 K)
of $\sim$490~meV and $\sim$580~meV for the c-ML and r-ML phases respectively.
Taking into account the difficulties in the calculations commented above, the 
agreement with the data in Table \ref{tab:The-interface-states} can be considered
quite good. 

Remarkably, and contrary to the experimental data, our
calculations predict the IS to be somewhat higher for the compressed layer, 
following the usually observed trend of the IS to upshift with increasing coverage. 
This trend is further confirmed by the computed data presented for the more dilute 
hypothetical 4$\times$4 phase. According to these data $\Delta E$ increases by 
$\sim$90~meV when the coverage increases from 0.672~ML (4$\times$4) to 
0.895~ML (r-ML), and  by additional $\sim$20~meV when increasing the coverage 
up to 1~ML for the c-ML phase.

Figure \ref{fig:Pbands} presents the corresponding band structures for the three studied structures. 
We identify an upshift of the IS with coverage to take place despite an enlargement of the vertical adsorption distances. 
Specifically, the functional group for the 4$\times$4 phase is located distinctly closer to the substrate 
($O_{carb}$: 2.53~\AA; $O_{anh}$: 2.77~\AA; $C_{func}$: 2.8~\AA)  in comparison with r-ML 
(see table \ref{tab:Rn-Vertical-adsorption-distance}). 
This should lead to a raising of the IS energy \cite{Zaitsev12prb}; 
however, this contribution apparently is noticeably weaker as compared to the effect of coverage.

The comparatively smaller increase of the IS energy when going from 
r-ML to c-ML is most likely due to the somewhat higher average positions of the molecules
in the c-ML phase, as discussed in conjunction with Fig.~\ref{fig:dz-distribution}. The larger distance
to the substrate thus tends to compensate the effect of the increasing coverage. 
Note, that the effective mass of the IS is about the same for all structures considered here 
($\sim 0.46~m_e$), and it is slightly bigger than the mass of the SS ($\sim 0.44~m_e$).

The competition between the effect of the molecular height 
and that of the coverage on the IS energy is explored in more detail 
in Table~\ref{tab:Interface-states-for-flat}. 
Here we used a flat NTCDA monolayer, for which the molecular
height over the silver surface $d_{\perp}$ can be defined unambigeously, 
and explore the dependence of the energy position of the IS for
molecular layers arranged both according to the r-ML and the c-ML phases. 
In this simplified model we neglect the bending of the molecules.

\begin{table}[tb]
\protect\caption{\label{tab:Interface-states-for-flat} 
Interface state energies for \emph{flat monolayers} of NTCDA on an Ag(111) slab with 12 silver layers. 
The relaxed and compressed structures with different adsorption heights of the flat monolayer are presented. 
The vertical bonding distances $d_{\perp}$ are given relative to the position of the topmost silver layer. 
The interface state energies for $d_{\perp}$ corresponding to experimental values are shown in bold.
}

\begin{ruledtabular}

\begin{tabular}{cccc}
$d_{\perp}$ [Å] & \multicolumn{3}{c}{$E_{IS}$ [meV]}\tabularnewline
\hline 
 & \multicolumn{1}{c}{relaxed} &  & \multicolumn{1}{c}{compressed}\tabularnewline
\cline{2-2} \cline{4-4} 
2.9 & 495  &  & 532\tabularnewline
3.0 & \textbf{428} &  & 447\tabularnewline
3.1 & 368 &  & \textbf{376}\tabularnewline
3.2 & 312 &  & 316\tabularnewline
3.3 & 265 &  & 267\tabularnewline
\end{tabular}

\end{ruledtabular}
\end{table}

We note that, even for the unrelaxed layers, the optimal height of the molecules over the Ag(111) substrate 
is close to that found for the carbon molecular framework for the fully optimized structures. 
Regarding the energy of the IS, for a given $d_{\perp}$ they are always higher for the compressed ML, 
especially for vertical distances $d_{\perp} \leq$ 3.1~\AA. 
Increasing $d_{\perp}$ to 3.2 Å and beyond, on the other hand, yields only minor differences. 
For a fixed $d_{\perp}$ (in the range 2.9 - 3.3~\AA), the effect of the larger coverage of c-ML 
produces shifts in the range from 2 to 37~meV. 

Interestingly, the data in  Table~\ref{tab:Interface-states-for-flat} tell that, 
comparatively, the effect of the molecular height on the IS position is
more important than that of coverage for NTCDA/Ag(111) and related systems. 
Over the full range of distances explored, a change of 0.1~\AA\ produces shifts of the IS energy ranging 
from 47 to 67~meV for the r-ML phase, and from 49 to 85~meV for the c-ML. 
Therefore, we can conclude that small height changes might easily compensate the effect of the coverage, 
and that the effect of the molecular height is somewhat larger for the c-ML phase. 
Measurements~\cite{Stanzel2004, Stadler2007} indicate that the 
vertical adsorption distance of the carbon atoms in the relaxed monolayer phase is 
close to 3.0~\AA\, while it is slightly lower than 3.1~\AA\ for the compressed phase. 
Taking those heights and using the corresponding data in  Table~\ref{tab:Interface-states-for-flat} 
(highlighted in bold format) we find that this height difference is sufficient to justify 
a $\sim$50~meV higher position for the IS of the r-ML phase, thus, largely 
compensating the effect of the coverage and bringing the calculated results 
in qualitative agreement with the experimental observation. 
The failure of the DFT approaches to fully account for the experimental findings 
(e.g. underestimating the difference in $d_{\perp}$ for c-ML and r-ML) 
is probably due to the limitations of the present functionals 
in describing the delicate interplay between the increased inter-molecular interactions 
in the compressed layer and the molecule-substrate interactions.

Therefore, our observation of a higher position of the IS for r-ML as compared to c-ML can be interpreted 
as an indication that the vertical separation for c-ML is notably larger than for r-ML. 
This causes a compensation of the upshift of $E_{IS}$ due to coverage. 
According to the experimental findings, which yield a value of $E_{IS}$ 
which is about 90~meV higher for r-ML as compared to c-ML, 
this compensation effect might be even larger than the estimates in the preceding paragraph.  
A much closer molecule-metal separation for r-ML is also consistent with the higher adsorption energy 
for this phase, as deduced from thermal desorption data \cite{braatz_vibrational_2012}.

\subsection{Summary}

For the model system NTCDA/Ag(111), 
the properties of the interface state (IS) of the relaxed 
and the compressed monolayer phases have been investigated 
using density-functional theory and two-photon photoemission. 

The dependence of the relaxed equilibrium configurations of the two NTCDA layers, 
and the corresponding energies of the IS, 
on two different approaches to include dispersive van der Waals interactions and 
two DFT implementations has been analyzed in detail. 
From a methodological point of view it is found that the energies of the 
surface state and IS are very dependent on the specifics of the calculations 
(e.g. through the different equilibrium lattice parameter with different computational approaches) 
and are very hard to converge (e.g. with respect to the number of layers in the slab). 
However, the errors in the energies of the surface state $E_{SS}$  and interface state $E_{IS}$
tend to cancel and the relative position of the IS with respect to the SS,  $\Delta E = E_{IS} - E_{SS}$,
is a more robust outcome of the calculation and provides the best way to compare against
experimental information, especially since $E_{SS}$ is known very accurately from the experiments. 

Two-photon-photoemission results provide accurate measurements of the energies of the 
IS, giving values of $\Delta E$ of $\sim$490~meV and $\sim$580~meV respectively 
for the  compressed and relaxed phases. These results are in reasonable agreement with 
the computed values. However, DFT calculations using the optimized geometries are unable to reproduce 
the counter-intuitive result that $\Delta E$ is higher for r-ML, in spite of its lower coverage $\Theta$.
DFT calculations predict that $\Delta E$ is roughly the same for both r-ML and c-ML, with 
$E_{IS}$ only $\sim$20~meV higher for the compressed layer. 
This small effect of the coverage difference can be understood from the higher average position 
of the molecules over the Ag(111) surface in the c-ML phase than in the r-ML. 
This also points to this height difference as the 
explanation for the experimental observation of a higher $E_{IS}$ for r-ML. 

In order to arrive at a fundamental and thorough understanding of this
interrelationship, and in an effort to bring together structural and spectroscopic information, 
the dependence of $E_{IS}$ with vertical bonding 
distances $d_{\perp}$ and the surface coverage $\Theta$ has been analyzed. 
While the coverage contributes substantially to determine the value of $\Delta E$ 
(especially for vertical bonding distances $d_{\perp} \leq$ 3.1~\AA), 
the impact of the molecular height  $d_{\perp}$ is even larger in the relevant range. 
Our calculations show that height differences in the order of $d_{\perp}$ = 0.1~\AA, i.e. consistent with the 
experimental information available for the geometries of r-ML and c-ML phases, are sufficient to compensate
the effect of the coverage difference and explain an $E_{IS}$ higher for r-ML than for c-ML. 

Given the fact that the energetic position of $E_{IS}$  for molecular layers on metal substrates 
depends crucially on various geometric as well as electronic properties (both of the molecule and the 
substrate surface layer) the found good agreement 
between the theoretically calculated interface energies 
and experimentally observed values is exceptional. 
This underlines the thorough understanding of molecule-metal interactions and the high accuracy 
that DFT methodologies are acquiring which, 
in particular, benefit from the inclusion of advanced van der Waals correction schemes.

\begin{acknowledgments}

This work is a project of the SFB 1083 "Structure and Dynamics of Internal Interfaces" funded by the Deutsche Forschungsgemeinschaft (DFG). DSP acknowledges support from the Spanish MINECO (Grant No. MAT2013-46593-C6-2-P). 

\end{acknowledgments}

\clearpage{}
\bibliographystyle{apsrev4-1}
\bibliography{lit}

\end{document}